\documentstyle[12pt,referee,psfig]{nature2}
\footnotesize
\def\deg{\ifmmode^{\circ}\else$^{\circ}$\fi}
\newdimen\minuswidth	
\setbox0=\hbox{$-$}
\minuswidth=\wd0
\newdimen\digitwidth	
\setbox0=\hbox{\rm0}
\digitwidth=\wd0
\catcode`!=\active
\def!{\kern\digitwidth}
\normalsize

\title{Transient radio bursts from rotating neutron stars}

\newcommand{\man}{\normalsize Jodrell Bank Observatory, University of Manchester, Macclesfield, Cheshire, SK11 9DL, UK}
\newcommand{\atnf}{\normalsize Australia Telescope National Facility -- CSIRO,
P.O. Box 76, Epping NSW 1710, Australia}
\newcommand{\cornell}{\normalsize Astronomy Department and NAIC, Cornell University, Ithaca, NY 14853, USA}
\newcommand{\columbia}{\normalsize Columbia Astrophysics Laboratory, Columbia University, 550 W. 120th Street,
New York, NY 10027, USA}
\newcommand{\cag}{\normalsize INAF - Osservatorio Astronomico di Cagliari, Loc. Poggio dei Pini,
Strada 54, 09012, Capoterra (CA), Italy}
\newcommand{\ubc}{\normalsize Dept. of Physics and Astronomy, University of British
Columbia, 6224 Agricultural Road, Vancouver, BC V6T 1Z1, Canada}
\newcommand{\nic}{\normalsize Dipartimento di Fisica, Universita di Cagliari, Strada Provinciale Monserrato-Sestu, km 0.700, 09042, Monserrato (CA), Italy}
\author{M.~A. McLaughlin\affiliation{\man}, A.~G.~Lyne$^{*}$, D.~R.~Lorimer$^{*}$,   M.~Kramer$^{*}$,  A.~J.~Faulkner$^{*}$,
R.~N.~Manchester\affiliation{\atnf},   J.~M.~Cordes\affiliation{\cornell}, F.~Camilo\affiliation{\columbia}, \\  A.~Possenti\affiliation{\cag},      I.~H.~Stairs\affiliation{\ubc},  G.~Hobbs$^{\dagger}$, N.~D'Amico\affiliation{\nic}$^{\P}$, \\
M. Burgay$^{\P}$ \& J.~T.~O'Brien$^{*}$\\}
\dates{}{}
\headertitle{}
\mainauthor{M.~A.~McLaughlin}
\begin{document}
\maketitle
{\bf \noindent The `radio sky' is relatively unexplored for transient
signals\cite{clm04}, although the potential of radio-transient
searches is high, as demonstrated recently by the discovery of a
previously unknown type of source\cite{hlk+05,kp05} which varies on
timescales of minutes to hours.  Here we report a new large-scale
search for radio sources varying on much shorter timescales. This has
revealed 11 objects characterized by single, dispersed bursts having
durations between 2 and 30~ms. The average time intervals between
bursts range from 4~minutes to 3~hours, with radio emission
typically detectable for ${\bf <1}$~s per day. From an analysis of the
burst arrival times, we have identified periodicities in the range
0.4--7 s for ten of the 11 sources, suggesting a rotating neutron star
origin. Despite the small number of sources presently detected, their
ephemeral nature implies a total Galactic population which
significantly exceeds that of the regularly pulsing radio pulsars.
Five of the ten sources have periods greater than 4~s, and period
derivatives have been measured for three of the sources, with one
having a very high inferred magnetic field of ${\bf 5\times10^{13}}$~G,
suggesting that this new population is related to other classes of
isolated neutron stars observed at X-ray and gamma-ray
wavelengths\cite{wt04b}.  }

The eleven sources were detected in a search for isolated bursts of
radio emission in data recorded for the Parkes Multibeam Pulsar Survey
between January 1998 and February 2002.  Figure~1 and 
Figure~2 show example detections.  All bursts from a given source have
the same unique value of dispersion measure (DM), or integrated
free-electron column density, incontrovertibly distinguishing them
from impulsive terrestrial interference.

Since August 2003, all the sources have been reobserved at least nine
times at intervals of between one and six months. All have shown
multiple bursts, with between four and 229 events detected in total
from each object (see Table~1).  As far as we can tell from the
limited statistics, the density of sources on the sky appears to be
greater towards the Galactic plane, with eight of the 11 having $|b| <
2\deg$.  Average rates of detected events range from one every three
hours for J1911+00 to one every 4 minutes for J1819--1458.  The
2~ms to 30~ms-long bursts have peak 1400-MHz flux densities which range from
0.1 to 3.6~Jy. These sources are therefore among the brightest radio
sources in the Universe after the giant pulses detected from the Crab
pulsar and the pulsar B1937+21\cite{cbh+04}.

Periodicity searches that depend on a pulsar's time-averaged emission,
including a standard Fourier analysis and a fast-folding
algorithm\cite{lk05}, have been carried out on all survey and
follow-up observations, with no periodicities detected using these
methods. However, for ten of the sources we have been able to identify
a periodicity from the arrival times of the bursts themselves (see
Table~2).  The 0.4 to 7~s period range indicates that they are likely
to be rotating neutron stars.  Most of the periods are quite long;
five of the 10 have periods exceeding four seconds, compared with only
$~1$ in 200 of the known radio pulsar population.  As shown in
Table~2, for three of the sources we have been able to measure period
derivatives using standard pulsar timing techniques\cite{lk05} on the
individual burst arrival times.

How is the bursting behaviour related to single-pulse behavior of
normal radio pulsars? For most of the sources, the number of detected
bursts is not yet sufficient to obtain reliable luminosity
distributions.  In Figure~3, we show the peak flux
density distributions for four objects.  For J1317--5759 and
J1819--1458, the periodic sources with the greatest number of bursts
detected, the non-detection in periodicity searches means that the
average peak flux density must be less than 0.5\% (for J1819--1458)
and 0.8\% (for J1317--5759) of the peak flux density of the strongest
detected bursts.  These objects show power-law tails to their burst
amplitude distributions, as seen for giant pulses from the Crab pulsar
and  pulsar B1937+21 (e.g. ref.~\pcite{cbh+04}).  However, all pulsars from
which giant pulses have been detected appear to have high values of
magnetic field strength at their light cylinder radii\cite{jr03}.
While the Crab pulsar has a magnetic field strength at the light
cylinder of $9.3\times10^{5}$~G, this value ranges from only 3 to 30~G
for these sources, suggesting that the bursts originate from a
different emission mechanism.

We therefore conclude that these sources represent a previously
unknown population of bursting neutron stars, which we call Rotating
RAdio Transients (RRATs).  In Figure~4, we show their relationship to
other neutron star populations. The long periods of some of the RRATs
are similar to the apparently radio-quiet X-ray populations of
magnetars\cite{wt04b} and isolated neutron stars\cite{hab04}.
Additionally, the inferred surface dipole magnetic field of
J1819--1458 of $5\times10^{13}$~G is greater than the magnetic field
of all but four of the 1600 known radio pulsars and is comparable to
those of the magnetars\cite{msk+03,wt04b}. This RRAT is young, with a
characteristic age of 117~kyr, smaller than those of 94\% of all
currently known radio pulsars.  The period and magnetic field of
J1317--5759 are similar to those of J0720.4--3125, the only radio-quiet isolated
neutron star with a measured period and period derivative\cite{kv05}.

The RRATs for which we have measured period
derivatives show no evidence for binary motion. Likewise, we detect no
glitches or other timing abnormalities, although continued monitoring
is necessary to gauge the regularity of spin-down rates. In the
future, radio polarization data may enable us to constrain the
emission mechanism.  For the three RRATs with accurate positions, a
search of high-energy archives\cite{heasarc} reveals no X-ray or
gamma-ray counterparts.

The discovery of this new population results in substantially
increased estimates of the total number of Galactic active
radio-emitting neutron stars. We detect, on average, one burst for
every three hours of observation for J1911+00. The chance of detection
within the single 35-min discovery observation was therefore less than
20\%, implying that there should be roughly five times the number of
similar sources in the same searched volume.  Applying a similar
analysis to all of the RRATs shows that we expect there to be twice as
many sources as we have detected at a similar sensitivity level and
sky coverage as for the Parkes survey.  This number may be a gross
underestimate, however. Firstly, it is very difficult to identify such
sources in observations which are contaminated with large amounts of
impulsive interference. There may be at least twice as many RRATs that
were missed due to this effect.  Secondly, we are only extrapolating
to the area covered by the Parkes survey, and the true distribution of
these objects is unknown.  In addition, because our sensitivity was
diminished for burst durations greater than 32~ms, there may be more
sources with longer bursts that fell below our detection threshold.
Furthermore, previous surveys with observation times of a few minutes
had little chance of detecting such events and most did not include
searches for them.

With these caveats in mind, we have carried out a Monte Carlo
simulation to provide a first-order estimate of the size of the
Galactic RRAT population. The simulation assumes that their spatial
distribution follows that derived for the pulsars detected in the
Parkes survey\cite{lor04}, that the burst-duration distribution is
similar to that observed in the 11 found so far, and, as measured for
the pulsar population\cite{lml+98}, that the differential radio
luminosity function of an average burst is of the form $d\log N/d\log
L=-1$, where $N$ is the number of model sources above a given
luminosity $L = Sd^2$, where $S$ is the peak flux density and $d$ is
the distance.  By calculating the threshold of our survey to model
bursts, and generating Monte Carlo realizations, we find the
simulations produce a good match to the observations, but are fairly
insensitive to the minimum burst peak luminosity $L_{\rm min}$, which
could plausibly lie in the range 1--100 mJy~kpc$^2$.  To be consistent
with the detection of 11 sources reported here, the implied size of
the Galactic population of RRATs $N \sim 4 \times 10^5 (L_{\rm min}/10
\, {\rm mJy \,\, kpc}^2)^{-1} \times (0.5/f_{\rm on}) \times
(0.5/f_{\rm int}) \times (0.1/f_b)$, where $f_{\rm on}$ is the
fraction of sources with bursts visible within our 35-min observation,
$f_{\rm int}$ is the fraction of bursts not missed due to interference
and $f_b$ is the fraction of RRATs whose bursts are beamed towards the
Earth. The average beaming fraction for pulsars is roughly 10\%, and
decreases for longer period pulsars\cite{tm98}. Given the small RRAT
duty cycles (see Table~2), our adopted $f_b$ is almost certainly a
conservative overestimate.  An $L_{\rm min}$ of 10~mJy~kpc$^{2}$ is
consistent with the lowest peak luminosities observed for the single
pulses of known radio pulsars.  Assuming that the total Galactic
population of active radio pulsars is of order $10^{5}$
(e.g.~ref.~\pcite{vml+04}), this discovery increases the current
Galactic population estimates by at least several times. We therefore
expect the emerging generation of wide-field radio telescopes to 
discover many more RRATs.


\vspace{0.2in}

\begin{acknowledge}
The Parkes radio telescope is part of the Australia Telescope which is
funded by the Commonwealth of Australia for operation as a National
Facility managed by CSIRO.  AP and ND'A acknowledge financial support
from the Italian Ministry of University and Research (MIUR).
FC is supported by NSF, NASA, and
NRAO. DRL is a University Research Fellow funded by the Royal Society.

\end{acknowledge}

\vspace{0.2in}

\noindent Reprints and permissions information is available at 
npg.nature.com/reprintsandpermissions.

\vspace{0.2in}

\noindent The authors declare that they have no competing financial interests.

\vspace{0.2in}

\noindent Correspondence and requests for materials should be
addressed to \verb+Maura.McLaughlin@manchester.ac.uk+.

\vspace{5in}

\begin{table*}[ht]
\begin{scriptsize}
\begin{center}
\begin{tabular}{lllcccccccc}
\\
\hline\hline Name & RA (J2000) & Dec (J2000) &  $l$  & $b$  & DM & $D$  & $w_{50}$  & $S_{1400}$ & $N_{p}$/$T_{obs}$ & $N_{det}$/$N_{obs}$  \\
& h~m~s & $\deg~'~''$ & $\deg$ & $\deg$ & pc cm$^{-3}$& kpc  & ms &  mJy  & hr$^{-1}$ & \\
\hline
J0848--43 & 08:48(1) & --43:16(7) & 263.4 & 0.2& 293(19) & 5.5 & 30 & 100 & 27/19 & 9/28 \\
J1317--5759 & 13:17:46.31(7) & --57:59:30.2(6) & 306.4 & 4.7 & 145.4(3) & 3.2  & 10 & 1100 &108/24&23/24 \\
J1443--60 & 14:43(1) & --60:32(7) & 316.2 & --0.6 & 369(8) & 5.5 & 20 & 280 & 32/41 & 17/25\\
J1754--30 & 17:54(1) & --30:11(7) & 359.9 & --2.2 & 98(6)  & 2.2 & 16 & 160 & 18/30 & 10/20\\
J1819--1458 & 18:19:33.0(5) & --14:58:16(32) & 16.0 & 0.1 & 196(3) &3.6& 3 & 3600 &229/13 & 24/24\\
J1826--14 & 18:26(1) & --14:27(7) & 17.2 & --1.0  & 159(1) & 3.3 & 2 & 600 & 18/17& 8/12\\
J1839--01 & 18:39(1) & --01:36(7) & 30.1 & 2.0  & 307(10) & 6.5& 15 & 100 & 8/13 & 1/10\\
J1846--02 & 18:46(1) & --02:56(7) & 29.7 & --0.1 & 239(10) & 5.2 & 16 & 250 & 11/10 & 5/9\\
J1848--12 & 18:48(1) & --12:47(7) & 21.1 & --5.0 & 88(2)& 2.4 & 2 & 450 & 10/8 & 5/9\\
J1911+00  & 19:11(1) & +00:37(7) & 35.7 & --4.1 & 100(3) & 3.3 & 5 & 250 & 4/13 & 4/11\\
J1913+1333  & 19:13:17.69(6) & +13:33:20.1(7) & 47.5 & 1.4 & 175.8(3) & 5.7 & 2 & 650 & 66/14 & 7/10\\
\hline
\end{tabular}
\end{center}
\label{tab:params}
\caption[]{{\bf Measured and derived parameters for the 11
sources.} For each, we give the Right Ascension, Declination, Galactic
longitude, Galactic latitude, DM, inferred distance, average burst
duration at 50\% of the maximum, peak 1400-MHz flux density of
brightest detected burst, ratio of the total number of bursts detected
to the total observation time, and the ratio of the number of
observations in which at least one burst was detected to the total
number of observations.  Estimated 1-$\sigma$ errors are given in
parentheses where relevant and refer to the last quoted digit. The
mean latitudes and longitudes are comparable to those of the pulsars
detected in the Parkes survey. The distances are inferred from their
DMs, positions and a model for the Galactic free electron
density\cite{cl02a}.  The mean distance of 4.2~kpc is comparable to
that of 5.8~kpc for the pulsars detected in the Parkes survey.  The
extremely sporadic nature of the bursts makes localization difficult,
with most positions known only to within the 1400-MHz 14-arcminute
beam of the Parkes Telescope.  For the three sources for which we have
measured period derivatives, more accurate positions have been derived
through radio timing.  Burst durations for each source remain
constant, within the uncertainties, and are all much larger than those
measured for pulsar giant pulses (e.g.~refs.~\pcite{spb+04},
\pcite{hkwe03}).
} 
\end{scriptsize}
\end{table*}

\begin{table*}[ht]
\begin{scriptsize}
\begin{center}
\begin{tabular}{llcccccc}
\\
\hline\hline
Name & $P$ & $w_{50}/P$ & Epoch & $\dot{P}$ & $B$ & $\tau_c$ & $\dot{E}$  \\
& s & \%  & MJD &  $10^{-15}$ s~s$^{-1}$ & $10^{12}$~G & Myr & $10^{31}$~erg~s$^{-1}$ \\
\hline
J0848--43  & 5.97748(2) & 0.50 & 53492 &  --  & -- & -- & -- \\
J1317--5759  &  2.6421979742(3) & 0.38 & 53346 & 12.6(7) & 5.83(2) & 3.33(2) & 2.69(1)  \\
J1443--60 &  4.758565(5)  & 0.42 & 53410 & -- & -- & -- & -- \\
J1754--30  &  0.422617(4) & 3.79 & 53189 &  -- & --  & -- & -- \\ 
J1819--1458  & 4.263159894(6) & 0.07 & 53265 & 576(1) & 50.16(6) & 0.1172(3) & 24.94(5) \\
J1826--14 & 0.7706187(3) & 0.26 & 53587 &   -- & -- & -- & --\\
J1839--01 & 0.93190(1) &  1.61 & 51038 & -- & -- &--  & -- \\
J1846--02  & 4.476739(3) & 0.36 & 53492 & -- & -- & -- & -- \\
J1848--12 & 6.7953(5) & 0.03 & 53158 & -- & -- & --  & -- \\
J1913+1333 & 0.9233885242(1) & 0.22 & 53264 &  7.87(2) & 2.727(4) & 1.860(6) & 39.4(1) \\
\hline
\end{tabular}
\end{center}
\label{tab:params2}
\caption{{\bf Measured and derived parameters of the 10
sources with measured periods.}  For each, we give the period, average
duty cycle (i.e. $w_{50}/P$), the epoch of the period and, if
measurable, the period derivative and derived parameters. Periods are
derived by calculating the largest common denominator of the
differences between the burst arrival times at a given epoch.  Given
the number of pulses detected per epoch, and the number of epochs for
which a periodicity can be measured, we may calculate the probability
that the listed period is an integer multiple of the true period.
Because of the small number of bursts detected per epoch for J1754--30
and J1848--12, there is a 32\% and 16\% chance, respectively, that the
listed period is actually an integer multiple of the true period.  For
J1839--01 and J1846--02, this probability is less than 1\% and for all
others, the probability is less than 0.1\%.
Assuming that they are rotating neutron stars, the inferred surface
dipole magnetic field is calculated as $B \equiv 3.2\times10^{19}
\sqrt{P\dot{P}}$~G, the characteristic age as $\tau_{c} \equiv
P/2\dot{P}$ and the spin-down luminosity as $\dot{E} \equiv
4\pi^2I\dot{P}P^{-3}$, where $I$, the neutron star moment of inertia,
is assumed to be $10^{45}$~g~cm$^{2}$ (see~ref.~\pcite{lk05}).  The
duty cycles are generally smaller than those of radio pulsars with
similar periods.
} 
\end{scriptsize}
\end{table*}

\begin{figure*}[t]
\center{\psfig{figure=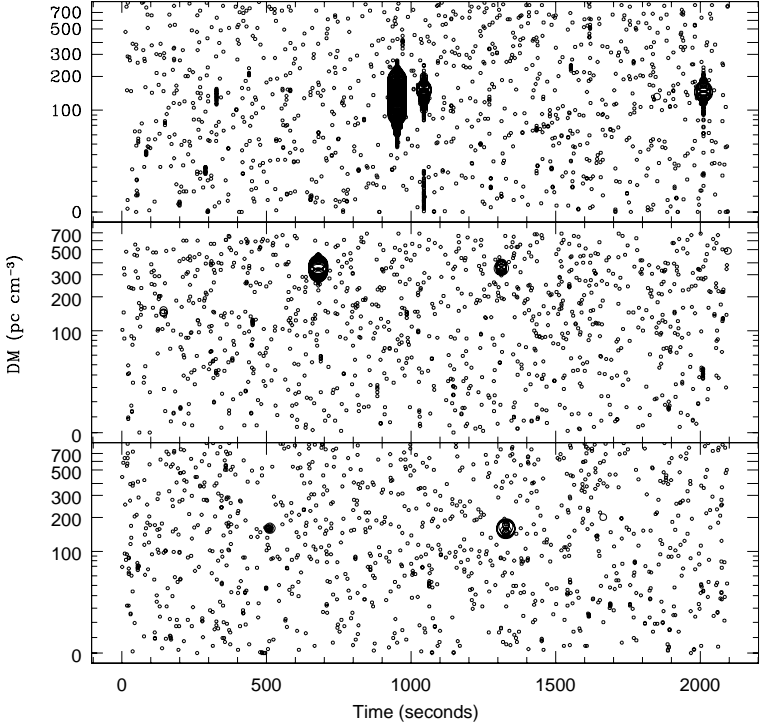,width=10cm}}
\caption[]{{\bf 
The observational signatures of the new radio
transient sources.}  From top to bottom, we show the original
detections of J1317--5759, J1443--60 and J1826--14 in the Parkes
Multibeam Survey data. The Parkes survey, which has discovered over
750 radio pulsars\cite{hfs+04}, used a 13-beam 1400-MHz cryogenic
receiver and covered 1500~deg$^{2}$ within 5$\deg$ of the Galactic
plane, for longitudes $260\deg < l < 50\deg$ with 250-$\mu$s sampling
of a multi-channel receiver and 35-min dwell-times on each
position\cite{mlc+01}.  Approximately 30\% of all pulsars that were
detected in the survey using standard periodicity-seeking Fourier
techniques were also detected in the burst search.  Since radio waves are
dispersed by ionised gas in the interstellar medium, the effects of
such dispersion have to be removed, and we have therefore used search
techniques similar to those described in ref.~\pcite{cm03}. In short,
the 35-minute time series were dedispersed for a number of trial
values of DM.  The time series were smoothed by convolution with
boxcars of various widths to increase sensitivity to broadened pulses,
with a maximum boxcar width of 32~ms. Because the optimal sensitivity
is achieved when the smoothing window width equals the burst width,
our sensitivity is lower for burst durations greater than 32~ms.  Each
of these time series was then searched for any bursts above a
threshold of five standard deviations, computed by calculating a
running mean and root-mean-square deviation of the noisy time series.
All bursts detected above a 5-$\sigma$ threshold are plotted as
circles, with size proportional to the signal-to-noise ratio of the
detected burst. The abcissa shows arrival time while the ordinate
shows the DM.  Because of their finite width, intense bursts are
detected at multiple DMs and result in vertical broadening of the
features. Bursts which are strongest at zero DM and therefore likely
to be impulsive terrestrial interference are not shown. In general
these were easily identified by their detection in multiple beams of
the 13-beam receiver.}
\end{figure*}

\begin{figure*}[t]
\center{\psfig{figure=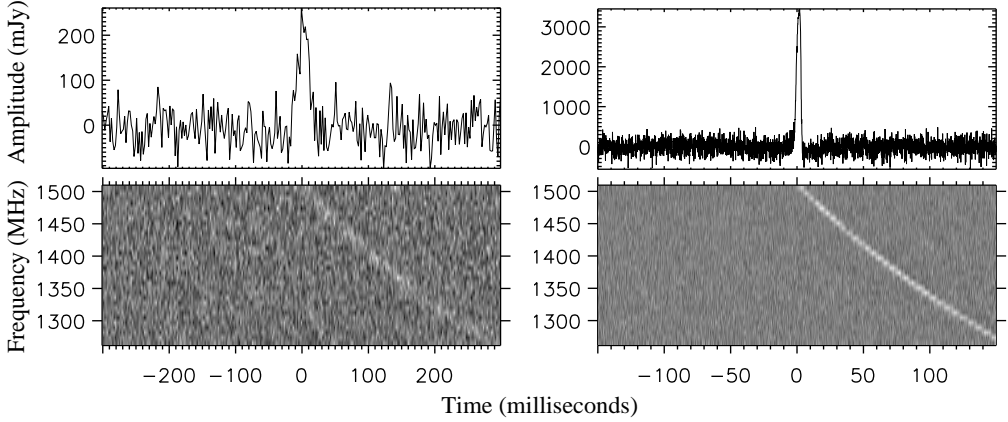,width=14cm}}
\caption[]{
{\bf Burst observational signatures in frequency
and time.}
The brightest single dispersed bursts detected from (left)
J1443--60 and (right) J1819--1458. The lower panel shows the
dispersed nature of the bursts detected in the individual frequency
channels.  The dispersion sweep is that expected for the radiation
from a celestial source after passing through the ionised gas of the
interstellar medium.  The upper panel shows the dedispersed time
series, obtained by summing outputs of the individual receiver
channels at the optimum value of the DM.
}
\end{figure*}

\begin{figure*}[t]
\center{\psfig{figure=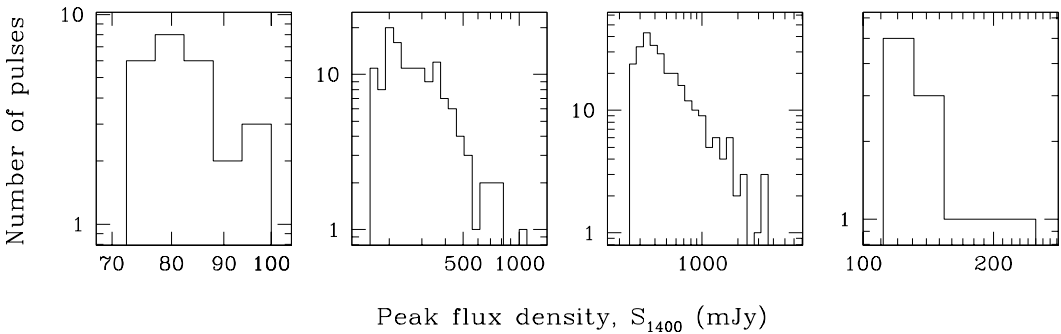,width=14cm}}
\caption[]{
{\bf Typical burst intensities.}
Histograms of the peak flux densities for (from left to
right): J0848--43, J1317--5759, J1819--1458 and J1846--02. The lower bound of all
histograms corresponds to a threshold of 6$\sigma$. The minimum detectable flux
density varies due to the
different burst widths. The pulse amplitude distributions are described by
power laws of index $\sim$ 1, less steep than the indices of 2 -- 3 measured for
giant pulsing
pulsars (e.g. refs.~\pcite{cm03}, \pcite{kt00}).}
\end{figure*}

\begin{figure*}[t]
\center{\psfig{figure=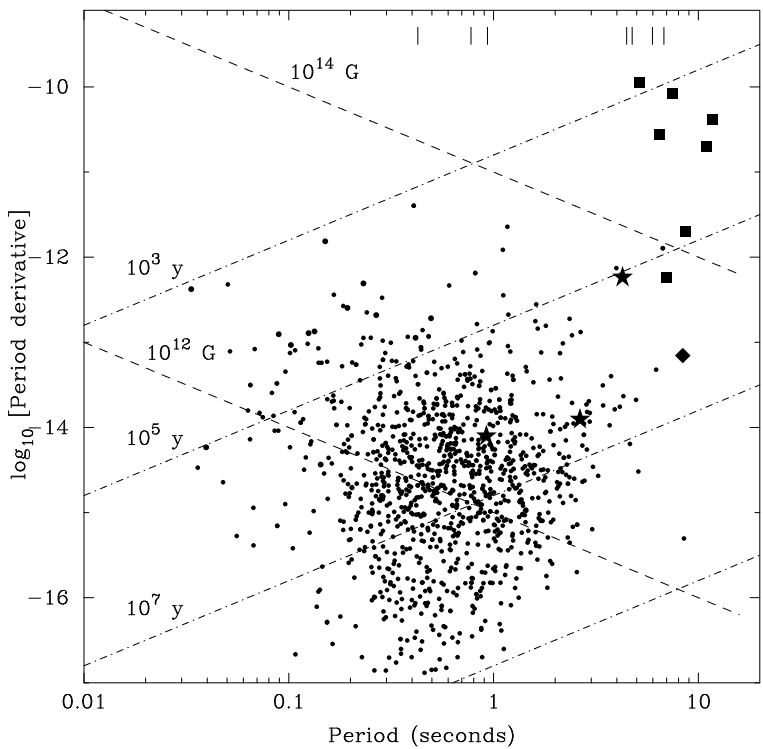,width=10cm,angle=270}}
\caption[]{
{
\bf The rotational properties of neutron stars
summarised in a {\bf $P-\dot{P}$} diagram.} The rotational period
derivative is plotted against period for pulsars (dots), magnetars
(squares), the one radio-quiet isolated neutron star with a measured period and
period derivative (diamond), and the three RRATs having measured
periods and period derivatives (stars). The vertical lines at the top
of the plot mark the periods of the other seven sources in
Table~2. Dashed lines indicate the loci of constant values of
characteristic age and inferred surface dipole magnetic field
strength.}
\end{figure*}

\end{document}